\documentstyle[11pt]{article}
\begin{document}

\title{Fock Space Structure for the Simplest Parasupersymmetric System
\thanks{This project supported by National Science Foundation of China and
LWTZ 1298.}}
\author{{Weimin Yang and Sicong Jing}\\
{\small Department of Modern Physics, University of Science and}\\
{\small Technology of China, Hefei 230026, P.R. China}}
\maketitle
\bigskip
\bigskip
\bigskip
\begin{abstract}
Structure of the state-vector space for a system consisting of one mode
parabose and one mode parafermi degree of freedom with the same parastatistics
order $p$ is studied and a complete, orthonormal set of basis vectors in this
space is constructed. There is an intrinsic double degeneracy for state
vectors with $m$ parabosons and $n$ parafermions, where $m \not= 0$,
$n \not= 0$, and $n \not= p$. It is also shown that the degeneracy
plays a key role in realization of exact supersymmetry for such a system.
\end{abstract}

\section{Introduction}
\par
One of the fundamental and powerful approaches in theoretical physics is a
search for symmetries. Generalized statistics first was introduced by Green in
the form of parastatistics as an exotic possibility extending the Bose and
Fermi statistics \cite{s1}. Recent research reveals that it probably has some
potential applications in the physics of the quantum Hall effect and the high
temperature superconductivity \cite{s2}. Supersymmetry, on the other hand,
unifies Bose and Fermi statistics and its development leads to important
progress in field and string theories \cite{s3}.
\par
A natural attempt is to construct a supersymmetric theory which can unify
parabose and parafermi statistics. Though some authors generalized the
supersymmetric quantum mechanics from the ordinary Bose and Fermi statistics
to parastatistics \cite{s4}, the so-called parasupersymmetric quantum
mechanics actually only unified ordinary bosons and parafermions, or ordinary
fermions and parabosons \cite{s5}. In other words, these works only unified 
parabosons and parafermions with different parastatistics order $p$. It is
well known from the parastatistics that if the parabosons and parafermions
are with different order $p$, they will commute or anticommute with each
other according to their algebraic relations being relative parabose or
relative parafermi, respectively \cite{s6}. In these cases the Fock space
structures of the parasupersymmetric systems are relatively simple. However,
if the parabosons and parafermions are with a same order $p$, they will
neither commute nor anticommute with each other and will satisfy some
complicated trilinear commutation relations. The Fock space structures for
these systems are still unknown yet, so far it is not really available to
construct a supersymetric quantum mechanics theory for parabosons and
parafermions with the same order $p$.
\par
In this paper, we first study the Fock space structure for a system with one
mode parabose and one mode parafermi degree of freedom which are with the same   
order $p$, and construct a complete, orthonormal set of basis vectors for this
parasystem. Due to the complexity of the parastatistics, there is an intrinsic
double degeneracy of state vectors in this Fock space. Based on this
understanding to the Fock space structure, we build a parasupersymmetric
quantum mechanics model for the same order $p$ parabosons and parafermions,
and the degeneracy plays an important role in the building.

\section{Fock space structure}
\par
It is well known that for one mode parabose system, the fundamental trilinear
relations are
\begin{equation}
\left[ a, \{ a^{\dagger}, a \} \right] = 2a, ~~~\left[ a, a^{\dagger 2}
\right] = 2a^{\dagger}, ~~~\left[ a, a^2 \right] = 0,
\end{equation}
where $a^{\dagger}$ and $a$ are parabose creation and annihilation operator
respectively, and for one mode parafermi system, the fundamental relation is
\begin{equation}
\left[ f, [f^{\dagger}, f] \right] = 2f,
\end{equation}
where $f^{\dagger}$ and $f$ are parafermi creation and annihilation operator
respectively. Moreover, for a system consisting of one parabose and one
parafermi degree of freedom with a same order $p$, besides the relations (1)
and (2), there are the following mixed trilinear relations \cite{s7}
\begin{eqnarray}
&& \left[ f, \{ a^{\dagger}, a \} \right] = 0, ~\left[ a, [f^{\dagger},f]
\right] = 0, ~\left[ f, a^2 \right] = 0, ~\left[ f^{\dagger}, a^2 \right] = 0,
\nonumber\\
&& \left[ a, \{ f^{\dagger}, a \} \right] = 0, ~\left[ a, \{ f, a \} \right]
= 0, ~\{ f, \{ f, a \} \} = 0, ~\{ f, \{ a^{\dagger}, f \} \} = 0, \nonumber\\
&& \left[ a, \{ f, a^{\dagger} \} \right] = 2f, ~\left[ a^{\dagger}, \{ a,
f \} \right] = -2f, ~\{ f, \{ f^{\dagger}, a \} \} = 2a, ~\{ f^{\dagger},
\{ a,f \} \} = 2a,
\end{eqnarray}
together with the adjoint ones. From (3) we see that one paraboson does
neither commute nor anticommute with parafermions, however, two parabosons
do commute with parafermions. According to parastatistics \cite{s8}, in
order to study the structure of a state vector space, we should impose the
following unique vacuum state conditions
\begin{eqnarray}
&& a|0 \rangle = 0, ~f|0 \rangle = 0, ~af^{\dagger}|0 \rangle = 0,
~fa^{\dagger}|0 \rangle = 0, \nonumber\\
&& aa^{\dagger}|0 \rangle = p|0 \rangle, ~ff^{\dagger}|0 \rangle =
p|0 \rangle.
\end{eqnarray}

\par
Let us denote the parabose and the parafermi number operator as $N_a$ and
$N_f$ respectively,
\begin{equation}
N_a = \frac{1}{2} \{a^{\dagger}, a\} - \frac{p}{2}, ~~~N_f = \frac{1}{2}
[f^{\dagger}, f] + \frac{p}{2}.
\end{equation}
Obviously, $N_a$ commutes with $N_f$, and the vacuum state $|0 \rangle$ is
their common eigenstate with the eigenvalue zero: $N_a |0 \rangle = 0$ and
$N_f |0 \rangle = 0$. For the sake of later convenience, we introduce four
useful operators $N_s$, $T$, $Q_s$ and $Q_s^{\dagger}$ which are defined by
\begin{eqnarray}
&& N_s = \frac{1}{p} \left( N_f^2 - (p+1) N_f + f^{\dagger}f + \frac{p}{2}
\right), \nonumber\\
&& T = \frac{p}{2} \left( F^{\dagger}F + Q^{\dagger}Q - N_a - \frac{p}{2}
\right) - 2 (N_a + \frac{p}{2})(N_f - \frac{p}{2})N_s, \nonumber\\
&& Q_s = (\frac{1}{2} - N_s) T, ~~~Q_s^{\dagger} = (\frac{1}{2} + N_s) T,
\end{eqnarray}
where $F = \frac{1}{2} \{a,f\}$ and $Q = \frac{1}{2} \{a^{\dagger}, f\}$. It
is readily to know the following relations between these operators
\begin{eqnarray}
&& [N_a, N_s] = 0, ~[N_f, N_s] = 0, ~[N_s, f] = 0, ~[N_s, a^2] = 0,
\nonumber\\
&& [N_a, T] = 0, ~[N_f, T] = 0, ~[N_a, Q_s] = 0, ~[N_f, Q_s] = 0, \nonumber\\
&& \{ N_s, T \} = 0, ~[ N_s, Q_s] = -Q_s, ~[Q_s^{\dagger}, Q_s] = 2T^2 N_s,
~\{ Q_s^{\dagger}, Q_s \} = T^2.
\end{eqnarray}

\par
We now investigate the state space structure for our parasystem. Firstly, we
search common eigenstates of the operators $N_a$ and $N_f$, which are denoted
by $|m,n \rangle$
\begin{equation}
N_a |m,n \rangle = m |m,n \rangle, ~~~N_f |m,n \rangle = n | m,n \rangle.
\end{equation}
Obviously, there are $m$ parabosons and $n$ parafermions in the state
$| m,n \rangle$. Using $[N_a, a^{\dagger n}] = n a^{\dagger n}$,
$[N_f, f^{\dagger n}] = n f^{\dagger n}$, $[N_a, f^{\dagger}] = 0$ and
$[N_f, a^{\dagger}] = 0$, we have

\par
{\bf Lemma 1}.  {\sl A state, produced by $m$ operators $a^{\dagger}$ and $n$
operators $f^{\dagger}$ acting on the vacuum state $|0 \rangle$ in any
permutation, is the common eigenstate of $N_a$ and $N_f$.}

\par
We can write the state $|m,n \rangle$ as
\begin{equation}
|m,n \rangle = f^{\dagger n_0}a^{\dagger m_1}f^{\dagger n_1}a^{\dagger m_2}
f^{\dagger n_2}...f^{\dagger n_{l-1}}a^{\dagger m_l}f^{\dagger n_l}|0 \rangle,
\end{equation}
where $m_1 + m_2 + ... + m_l = m$, $n_0 + n_1 + n_2 + ... + n_l =n$, and
all the $m_i \geq 1$, $(i=1,...,l)$, all the $n_i \geq 1$, $(i=1,...,l-1)$,
with $n_0, n_l \geq 0$. From the Lemma 1 we see that any two partitions
$(m_1, m_2, ..., m_l)$ and $(n_0, n_1, n_2, ..., n_l)$ satisfying
$\sum_{i=1}^{l} m_i = m$, $\sum_{i=0}^{l} n_i = n$ will correspond to a state
$|m,n \rangle$. Since $a^{\dagger}$ and $f^{\dagger}$ neither commute nor
anticommute with each other, it is quite possible that several different
permutations of the operators $a^{\dagger}$ and $f^{\dagger}$ correspond to a
same state $|m,n \rangle $, or that there is an intrinsic degeneracy in the
subspace spanned by different eigenstates of $N_a$ and $N_f$ with the same
eigenvalues $m$ and $n$ (we use the symbol (m,n) to denote this subspace). In
order to make the structure of the subspace (m,n) clear, we furthermore have

\par
{\bf Lemma 2}.  {\sl The subspace (m,n) is usually two dimensional, except for three
special cases, i.e., for (0,n), (m,0) and (m,p), in which the intrinsic
degeneracy disappears.}

\par
To prove the Lemma 2, let us use $[a^{\dagger 2}, f^{\dagger}] = 0$, and shift
all the even powers of $a^{\dagger}$ in the right-hand of (9) to left side
of the permutation of $a^{\dagger}$ and $f^{\dagger}$, which will lead to
\begin{equation}
|m,n \rangle = a^{\dagger (m-s)} f^{\dagger k_0}a^{\dagger}f^{\dagger k_1}
a^{\dagger}f^{\dagger k_2}...f^{\dagger k_{s-1}}a^{\dagger}f^{\dagger k_s}
|0 \rangle,
\end{equation}
where $m-s$ is an even integer, and $k_0 + k_1 + k_2 + ... + k_s = n$ with
$k_1, k_2, ..., k_{s-1} \geq 1$ and $k_0, k_s \geq 0$. Then repeatedly using
the relations $a^{\dagger}f^{\dagger k} = (-)^k f^{\dagger k}a^{\dagger}
+ 2k F^{\dagger}f^{\dagger (k-1)}$, $[F^{\dagger}, a^{\dagger}] = 0$,
$\{ F^{\dagger}, f^{\dagger} \} = 0$ and $F^{\dagger 2} = 0$, we may rewrite
the permutation $a^{\dagger}f^{\dagger k_1}a^{\dagger}f^{\dagger k_2}...
a^{\dagger}f^{\dagger k_s}$ as a sum of two terms
\begin{equation}
a^{\dagger}f^{\dagger k_1}a^{\dagger}f^{\dagger k_2}...
a^{\dagger}f^{\dagger k_s} = \alpha_s f^{\dagger (k_1 + k_2 + ... + k_s)}
a^{\dagger s} + \beta_s f^{\dagger (k_1 + k_2 + ... + k_s - 1)}
a^{\dagger (s-1)} F^{\dagger},
\end{equation}
where $\alpha_s$ and $\beta_s$ are two coefficients decided by the integer
$s$. For instance, we have
\begin{eqnarray*}
&& a^{\dagger}f^{\dagger k_1} = (-)^{k_1}f^{\dagger k_1}a^{\dagger}
+ (-)^{k_1 -1} 2k_1 f^{\dagger (k_1 -1)}F^{\dagger}, \nonumber\\
&& a^{\dagger}f^{\dagger k_1}a^{\dagger}f^{\dagger k_2} =
(-)^{k_1} f^{\dagger (k_1 + k_2)}a^{\dagger 2} + (-)^{k_1 - 1} 2k_1
f^{\dagger (k_1 + k_2 -1)}a^{\dagger}F^{\dagger}, \nonumber\\
&& a^{\dagger}f^{\dagger k_1}a^{\dagger}f^{\dagger k_2}a^{\dagger}
f^{\dagger k_3} = (-)^{k_1 + k_3} f^{\dagger (k_1 + k_2 + k_3)}a^{\dagger 3}
+ (-)^{k_2} 2(k_1 + k_3) f^{\dagger (k_1 + k_2 + k_3 -1)}a^{\dagger 2}
F^{\dagger},
\end{eqnarray*}
and so on. Substituting (11) into (10), we arrive at
\begin{equation}
|m,n \rangle = \alpha_s f^{\dagger n}a^{\dagger m} |0 \rangle +
\beta_s f^{\dagger (n-1)}a^{\dagger (m-1)}F^{\dagger} |0 \rangle,
\end{equation}
which means that we can write any state $|m,n \rangle$ (9) satisfying the
conditions $\sum_{i=1}^l m_i = m$ and $\sum_{i=0}^l n_i =n$ as a linear
combination of two basic state-vectors $f^{\dagger n}a^{\dagger m} |0 \rangle$
and $f^{\dagger (n-1)}a^{\dagger (m-1)}F^{\dagger} |0 \rangle$. It is worth
pointing out that for the $n \geq p+1$ case, both the two basic state-vectors
will become zero because the parastatistics order is $p$. Thus the state
$|m,n \rangle$ will vanish for $n \geq p+1$ case. As for why the degeneracy in
the subspace (m,n) will disappear for three special case (i.e., for (0,n),
(m,0) and (m,p) cases), please see the words following (15). Combining the
Lemma 1 and the Lemma 2, we have the following theorem

\par
{\bf Theorem}. {\sl Any common eigenstate $|m,n \rangle$ of the parabose nnumber
operator $N_a$ and the parafermi number operator $N_f$ with the eigenvalues
$m$ and $n$ respectively can be expressed as a combination of two basic
state-vectors $f^{\dagger n}a^{\dagger m}|0 \rangle$ and $f^{\dagger (n-1)}
a^{\dagger (m-1)}F^{\dagger}|0 \rangle$, as in (12), where $m \not= 0$,
$n \not= 0$ and $n \not= p$.}

\par
We now denote the two basic state-vectors as $|m,n,\alpha \rangle =
f^{\dagger n}a^{\dagger m}|0 \rangle$ and $|m,n,\beta \rangle =
f^{\dagger (n-1)}a^{\dagger (m-1)}F^{\dagger}|0 \rangle$. After some algebras
we see that usually the inner products of the two basic state-vectors are not
zero, i.e., $\langle m,n, \alpha |m,n, \beta \rangle \not= 0$, or the two
basic state-vectors are not orthogonal. Using Schmidt method we can get two
orthonormal basis vectors in the subspace $(m,n)$
\begin{eqnarray}
&& |m,n, s=\frac{1}{2} \rangle = \sqrt{\frac{(p-n)!}{p!n![m]!}} f^{\dagger n}
a^{\dagger m}|0 \rangle, \nonumber\\
&& |m,n, s=-\frac{1}{2} \rangle = \sqrt{\frac{(p-n-1)!(m+\frac{1-(-)^m}{2})}
{p!(n-1)![m+1]!}}f^{\dagger (n-1)} \left( pF^{\dagger} -
f^{\dagger}a^{\dagger} \right)a^{\dagger (m-1)}|0 \rangle,
\end{eqnarray}
which satisfy
\begin{equation}
\langle m,n,s |m^{\prime},n^{\prime},s^{\prime} \rangle =
\delta_{m,m^{\prime}} \delta_{n,n^{\prime}} \delta_{s,s^{\prime}},
\end{equation}
and
\begin{equation}
\sum_{m=0}^{\infty} \sum_{n=0}^{p} \sum_{s=\pm 1/2} |m,n,s \rangle
\langle m,n,s| = 1,
\end{equation}
where $[m] = m + (p-1)\frac{1-(-)^m}{2}$, $[m]! = [m][m-1]...[1]$ and
$[0]! = 1$. Some remarks concerning the basis vector $|m,n,s \rangle$ are as
follows. Firstly, for $m=0$ or $n=0$ case, by the construction of the state
$|m,n \rangle$, only the state $|m,n,1/2 \rangle$ exists, which exactly
coincides with the basis vectors of one mode parafermi system or the basis
vectors of one mode parabose system respectively. This can also be seen from
the normalization factor of the state $|m,n,-1/2 \rangle$. For example, when
$m=0$, the normalization factor of the state $|m,n,-1/2 \rangle$ is zero,
which means the state $|m,n,-1/2 \rangle$ does not exist. Secondly, for $n=p$
case, using a formula $pf^{\dagger (p-1)}a^{\dagger}f^{\dagger} +
(p-2)f^{\dagger p}a^{\dagger} = 0$ which works for paraboson and parafermion
with the same order $p$, we also see the state $|m,n,-1/2 \rangle$ vanishes.
So in the three special cases (m=0, n=0, or n=p), there exist only the states
$|m,n,1/2 \rangle$, or the intrinsic double degeneracy in the subspace (m,n)
disappears. Needless to say, if the order $p=1$, our parasystem will reduce
to the ordinary boson and fermion system, and of course the degeneracy will
disappear. Also when $n \geq p+1$, all the basis vectors will become zero.
\par
Furthermore, we show that the two basis vectors $|m,n,1/2 \rangle$ and
$|m,n,-1/2 \rangle$ are eigenstates of the operator $N_s$ defined by (6) with
eigenvalue $1/2$ and $-1/2$ respectively. In fact, noticing
$[N_s, f^{\dagger}] = 0$, $[N_s, a^{\dagger 2}] = 0$, and
$N_s|0 \rangle = \frac{1}{2}|0 \rangle$,
$N_s a^{\dagger}|0 \rangle = \frac{1}{2} a^{\dagger}|0 \rangle$, we have
\begin{equation}
N_s|m,n,1/2 \rangle = \frac{1}{2}\,|m,n,1/2 \rangle.
\end{equation}
Then using $N_f F^{\dagger}|0 \rangle = F^{\dagger}|0 \rangle$, and
$N_f F^{\dagger}a^{\dagger}|0 \rangle = F^{\dagger}a^{\dagger}|0 \rangle$, we
have
\begin{eqnarray}
&& N_s (pF^{\dagger} - f^{\dagger}a^{\dagger})|0 \rangle =
-\frac{1}{2} (pF^{\dagger} - f^{\dagger}a^{\dagger})|0 \rangle, \nonumber\\
&& N_s (pF^{\dagger} - f^{\dagger}a^{\dagger}) a^{\dagger}|0 \rangle =
-\frac{1}{2} (pF^{\dagger} - f^{\dagger}a^{\dagger}) a^{\dagger}|0 \rangle,
\end{eqnarray}
which will lead to
\begin{equation}
N_s |m,n,-1/2 \rangle = -\frac{1}{2}\,|m,n,-1/2 \rangle.
\end{equation}
Thus we clearly understand the Fock space structure of our parasystem. The set
of complete and orthonormal basis vectors in this Fork space is
$\{ |m,n,s \rangle \}$ which are common eigenstates of three operators $N_a$,
$N_f$ and $N_s$.
\par
Before concluding this section, let us see actions of the operators $T$, $Q_s$
and $Q_s^{\dagger}$ defined by (6) on the basis states $|m,n,s \rangle$. After
some simple calculations we find that
\begin{eqnarray}
&& T|m,n,1/2 \rangle = \sqrt{n(p-n) \left( m^2 + mp + p^2(1-(-)^m)/8 \right)}
|m,n,-1/2 \rangle,  \nonumber\\
&& T|m,n,-1/2 \rangle = \sqrt{n(p-n) \left( m^2 + mp + p^2(1-(-)^m)/8 \right)}
|m,n,1/2 \rangle,
\end{eqnarray}
which mean that $T$ is a transition operator changing $|m,n,1/2 \rangle$ to
$|m,n,-1/2 \rangle$ and vice versa. In addition, we also have
\begin{eqnarray}
&& Q_s|m,n,1/2 \rangle = \sqrt{n(p-n)\left( m^2 + mp + p^2(1-(-)^m)/8 \right)}
|m,n,-1/2 \rangle, \nonumber\\
&& Q_s|m,n,-1/2 \rangle = 0,
\end{eqnarray}
and
\begin{eqnarray}
&& Q_s^{\dagger}|m,n,1/2 \rangle = 0, \nonumber\\
&& Q_s^{\dagger}|m,n,-1/2 \rangle = \sqrt{n(p-n)\left( m^2 +mp +
p^2(1-(-)^m)/8 \right)}|m,n,1/2 \rangle,
\end{eqnarray}
which imply that $Q_s$ and $Q_s^{\dagger}$ are ladder operators in the two
dimensional subspace (m,n).

\section{Parasupersymmetric quantum mechanics model}
\par
Now we consider the simplest parasupersymmetric model consisting of a free
parabose oscillator and a free parafermi oscillator with the Hamiltonian
\begin{equation}
H = \frac{1}{2} \{ a^{\dagger}, a\} + \frac{1}{2} [f^{\dagger}, f]
= N_a + N_f.
\end{equation}
>From our basic relations in (3), we know that the operators $H$, $Q$ and
$Q_{\dagger}$ satisfy the standard supersymmetric algebra
\begin{equation}
[H,Q] = [H, Q^{\dagger}] = 0, ~~~\{ Q, Q^{\dagger} \} = H,
~~~Q^2 = Q^{\dagger 2} = 0,
\end{equation}
so $Q$ and $Q^{\dagger}$ are aupersymmetric charge operators which realize the
supersymmetry between parabosons and parafermions. It is easily to see that
$Q$ (or $Q^{\dagger}$) changes states between two subspaces (m,n) and
(m+1,n-1) (or (m,n) and (m-1,n+1)). In the previous section we construct a set
of complete and orthonormal basis vectors $\{ |m,n,s \rangle \}$ for our
parasystem. In this basis the spectrum of the Hamiltonian (22) can be
immediately obtained
\begin{equation}
H|m,n,s \rangle = E_{m+n}|m,n,s \rangle = (m+n)|m,n,s \rangle,
\end{equation}
where $m \geq 0$ and $0 \leq n \leq p$. We would like to emphasize here that
there is an intrinsic double degeneracy for any state with $m$ parabosons and
$n$ parafermions, except for three extreme cases in which $m=0$, or $n=0$, or
$n=p$. For $p=3$ case, the spectrum is shown in Fig.1. From the figure we see
that the ground state level $E_0 =0$ is non-degenerate, which results from the
assumption of the unique vacuum state and also means that the
parasupersymmetry is unbroken. The first level $E_1$ is double degenerate, to
which there are two states $|1,0,1/2 \rangle$ and $|0,1,1/2 \rangle$
corresponding.The second level $E_2$ is quadruple degenerate, to which four
states $|2,0,1/2 \rangle$, $|0,2,1/2 \rangle$, $|1,1,1/2 \rangle$ and
$|1,1,-1/2 \rangle$ correspond. Generally speaking, the degeneracy of the
$l$-th level $(l\leq p)$ $E_l$ is $2l$-fold, or there are $2l$ states with the
energy $E_l = l$, which are $|l,0,1/2 \rangle$, $|0,l,1/2 \rangle$, and
$|m,n,\pm 1/2 \rangle$, where $m=1,2,...,l-1$, so $n=l-1,l-2,...,1$,
respectively. However, for $l>p$ case, the degeneracy of the level $E_l$ is   
the same $2p$-fold, or there are always $2p$ states having the energy
$E_l = l$, which are $|l,0,1/2 \rangle$, $|l-p,p,1/2 \rangle$, and
$|m,n,\pm 1/2\rangle$, where $n=1,2,...,p-1$ and correspodingly
$m=l-1,l-2,...,l-p+1$.
\par
Here we find that the intrinsic double degeneracy in the subspace (m,n) plays
an important role in realization of the supersymmetry for our parasystem. From
the ordinary supersymmetric quantum mechanics we know that the number of bose
states is equal to the number of fermi states in the Fock space of the theory
\cite{s9}. If one introduces the Witten index \cite{s10}, or an operator
$(-)^{N_F}$, where $N_F$ is the fermi number operator, then one must have
$Tr\,(-)^{N_F} = 0$ for a supersymmetric quantum system, here the trace is
done for the all physical states except the vacuum state. In our case, we
should generalize the Witten index to the form of $(-)^{N_f}$, where the
$N_f$ is the parafermi number operator, and if our parasystem is
supersymmetry, we must also have $Tr\,(-)^{N_f} = 0$, here the trace is done
for our all states except
the vacuum state. In other words, for our parasystem, if the supersymmtry is
exactly established, the number of states with even parafermions must exactly
be equal to the number of states with odd parafermions. Due to the intrinsic
double degeneracy in the subspace (m,n) (where $m \not= 0$, $n \not= 0$ and
$n \not= p$), $Tr\,(-)^{N_f}$ is exact zero, here the trace is done for all
states except the vacuum state. To see this, let us consider every excitation
level $E_l$. In every such level, we have even states no matter
$l \leq p$ or $l>p$, and half of them are even- and half of them are
odd-parafermion states, so $Tr\,(-)^{N_f} = 0$. If we had without the double
degeneracy, we would have no way to realize the supersymmetry between the
parabosons and the parafermions. It is exactly in this sense, we say that the
intrinsic double degeneracy plays an fundamental rule in building the
parasupersymmetry for our system.

\section{Conclusion}
\par
In this paper, the Fork space structure of a system consisting of one mode
parabose and one mode parafermi degree of freedom with the same parastatistics
order $p$ is investigated. Since the parabosons neither conmmute nor
anticonmmute with the parafermions, the Fork space of our parasystem is more
complicated than one of the ordinary bose and fermi systems, or in other
words, the Fork space of our parasystem is not simply a direct product of
parabose subspace and parafermi subspace. It is fortunately that there is a
double degeneracy in the subspace (m,n) which insures that the number of
even-parafermion states is exactly the same as the number of odd-parafermion
states. Based on this, the unbroken supersymmetry is established for our
parasystem. An interesting task is to generalize the result of this paper to
more complicated parasupersymmetric systems, and work on this direction is in
progress.

\end{document}